\documentclass[twocolumn]{aastex62}
\def\secref#1{Section~\ref{#1}}
\begin{document}

\title{Measuring and modeling the rate of separator reconnection between an emerging and existing active region}
\author{Marika I. McCarthy}
\affiliation{Department of Physics, Montana State University, Bozeman, MT 59717}

\author{Dana W. Longcope}
\affiliation{Department of Physics, Montana State University, Bozeman, MT 59717}

\author{Anna Malanushenko}
\affiliation{High Altitude Observatory, National Center for Atmospheric Research, Boulder, CO 80301}

\author{David E. McKenzie}
\affiliation{NASA Marshall Space Flight Center, Huntsville, AL 35812}

\begin{abstract}
Magnetic reconnection occurs when new flux emerges into the corona and becomes incorporated into the existing coronal field.  A new active region (AR) emerging in the vicinity of an existing AR provides a convenient laboratory in which reconnection of this kind can be quantified.  We use high time-cadence 171 $\AA$ data from SDO/AIA focused on new/old active region pair 11147/11149, to quantify reconnection.  We identify new loops as brightenings within a strip of pixels between the regions.  This strategy is premised on the assumption that the energy brightening a loop originates in magnetic reconnection.  We catalog 301 loops observed in the 48-hour time period beginning with the emergence of AR 11149.  The rate at which these loops appear between the two ARs is used to calculate the reconnection rate between them.  We then fit these loops with magnetic field, solving for each loop's field strength, geometry, and twist (via its proxy, coronal $\alpha$).  We find the rate of newly-brightened flux overestimates the flux which could be undergoing reconnection.  This excess can be explained by our finding that the interconnecting region is not at its lowest energy (constant-$\alpha$) state; the extrapolations exhibit loop-to-loop variation in $\alpha$.  This flux overestimate may result from the slow emergence of AR 11149, allowing time for Taylor relaxation internal to the domain of the reconnected flux to bring the $\alpha$ distribution towards a single value which provides another mechanism for brightening loops after they are first created.
\end{abstract}

\section{Introduction}
  
The solar corona is a dynamic region where both the coronal magnetic field and the plasma which traces it is constantly evolving.  Complicated magnetic fields can lead to energetic events, particularly in regions of strong magnetic fields such as active regions (ARs), where the Sun's magnetic field breaches the surface.  The mechanism believed to drive events like flares \citep{1958NCim....8S.188S}, or the heating of the corona \citep{1972ApJ...174..499P} is magnetic reconnection.  In most theories of reconnection, energy is released as magnetic field line connectivity is rearranged \citep{1958NCim....8S.188S,PhysRev.107.830,1964NASSP..50..425P}.  While there are many theories about how this process takes place, there are not so many quantifying observations.

Previous studies looked to ARs to provide observational evidence of magnetic reconnection. In several studies, the formation of coronal loops between two different ARs has been used as evidence of nonflaring reconnection in the corona \citep{1996ApJ...456L..63T,1975SoPh...40..103S,Webb1981}.  Work done by \cite{Tarr2014} examined quiescent reconnection within single NOAA AR11112 and inferred a reconnection rate therein.  Analysis of a single AR, however, can be difficult to interpret due to the ambiguity in distinguishing between old and new photospheric flux. 

The emergence of a new AR in the vicinity of an existing one provides a good laboratory in which to study and quantify magnetic reconnection.  Under the prevailing understanding, each AR originates as an isolated magnetic flux tube \citep{1994ApJ...436..907F,2000SoPh..192..119F,2001ApJ...559L..55M}.  This flux tube is buoyant, rises and breaches the surface, and expands to fill a larger volume in the corona.  Two ARs that are adjacent to each other will be distinct systems, yet expand into contact in the corona.  
Under this assumption, any coronal loops that interconnect the two ARs must be formed by magnetic reconnection; this is the process that allows the connectivities to change.  This has been modeled in various numerical magnetohydrodyamic (MHD) simulations of emerging flux interacting with other fields in a model corona \citep{Galsgaard_2007,2011A&A...531A.108M,2017ApJ...850...39T}.  The slower emergence, compared to impulsive reconnection in a flare, provides an extended opportunity to observe and make measurements of any reconnection that occurs during the event.

\cite{2005ApJ...630..596L} performed this sort of two-active-region analysis with the {\it Transition Region and Coronal Explorer} \citep[TRACE;][]{Handy1999}.  They studied a single pair of active regions, and found no other suitable candidates for analysis in the TRACE archives.  The limited FOV of that instrument required a deliberate pointing for an extended period of time to observe an emergence and the subsequent magnetic evolution.  \cite{2005ApJ...630..596L} reported that reconnection, measured using the newly formed loops observed between the ARs, did not occur at the rate they inferred from the magnetic evolution of the system.  There was instead a delay of approximately 24 hours between emergence of the new AR and its reconnection to the overlying field based on loop measurements, while the magnetic modeling was suggestive of no delay between emergence and the onset of reconnection.  An observation by \cite{2008A&A...488.1117Z} has also noted a delay of $\sim$12 hour between flux emergence within an existing AR and the first appearance of coronal loops between the old and new flux.  \cite{KobelskiThesis} used SWAP to look at 8 flux emergence events and also observed a delay between emergence and coronal loops as evidence of reconnection to a nearby pre-existing AR.  Those observed delays generally were around 25 hours.  However, limitations on the temporal and spatial resolution of SWAP lead to the interpretation of these values as an upper bound for the time delay.  More observations are required before it is possible to further characterize the specific properties of flux emergence and subsequent reconnection.

The Atmospheric Imaging Assembly (AIA) on the Solar Dynamics Observatory \citep[SDO;][]{2012SoPh..275...17L} observes the full solar disk every $12$s thus providing continuous, high-cadence observation of active region emergence.  With AIA we need not wait for a serendipitous instrument pointing to ``catch'' this kind of emergence event, nor do we need to monopolize the instrument as the event occurs.  As such, this mission offers many more candidates for analysis.  This is ideal for elucidating how magnetic reconnection is involved when two ARs come into contact in the corona.

The present work analyzes data from 2011 January 20-22, wherein NOAA AR11149 emerged to the south of AR11147 as the two AR system crossed disk center in the northern hemisphere.  We ultimately find that the reconnected flux inferred from the appearing coronal loops is comparable to the total flux in either AR, overestimating the reconnected portion of its flux.  In order to explain this we argue that a single flux element in the reconnected domain appears more than once, manifesting as re-brightening coronal loops.  This interpretation constitutes a significant difference between the new case and that analyzed in \cite{2005ApJ...630..596L}.  It is necessary to explore and understand the different scenarios represented by these two cases if we are to ultimately form a general picture of coronal reconnection between ARs.  In order to obtain the best estimates of reconnected flux, the present work replaces potential-field extrapolation with a linear force-free model of individual loops in the coronal field.   The improved techniques detailed herein will later be applied in a larger set of $\mathbf{17}$ emerging/existing AR pairs to quantify reconnection between them.

We describe the analysis in the following sections.
We build a potential field model from SDO/HMI data in \secref{sec:MCT}, to produce a context calculation illustrating the separator reconnection process in which we are interested.  \secref{sec:data} uses data from SDO/AIA to construct a time/space stack plot from which we determine the reconnection rate, computed from the identified interconnecting loops.  The process of cataloging the loops is explained in that section and their parameters are reported.  An improved method for determining magnetic field's structure and properties is detailed in \secref{sec:annymodel}.   Loops are observed in plane-of-sky (POS) and their POS projections are 2D curves.  The POS curve of a coronal loop previously identified in the data is fit to a three-dimensional field line in a linear force-free field (LFFF) model.  As a result, we obtain a three-dimensional model of that loop.These results are used in \secref{sec:nonzeroalpha} to reexamine both the reconnection rate and the potential field context calculation.  Reasons for discrepancies are then explored.  
With the benefit of the 3D tracks of coronal loops obtained through the LFFF modeling, in \secref{sec:slitplane} we are able to use the line-of-sight (LOS) coordinate to show where the loops lie in a plane defined by the LOS and the pixels from which the stack plot was built.
We find there is significant overlap of the catalog loops' cross-sections in that plane.  Various rates of brightening are explored in that section before discussion in \secref{sec:dis} regarding possible rebrightening of magnetic flux tubes inside of the reconnected flux domain.

\section{Potential field model} \label{sec:MCT}

\begin{figure}[]
\centering
\plotone{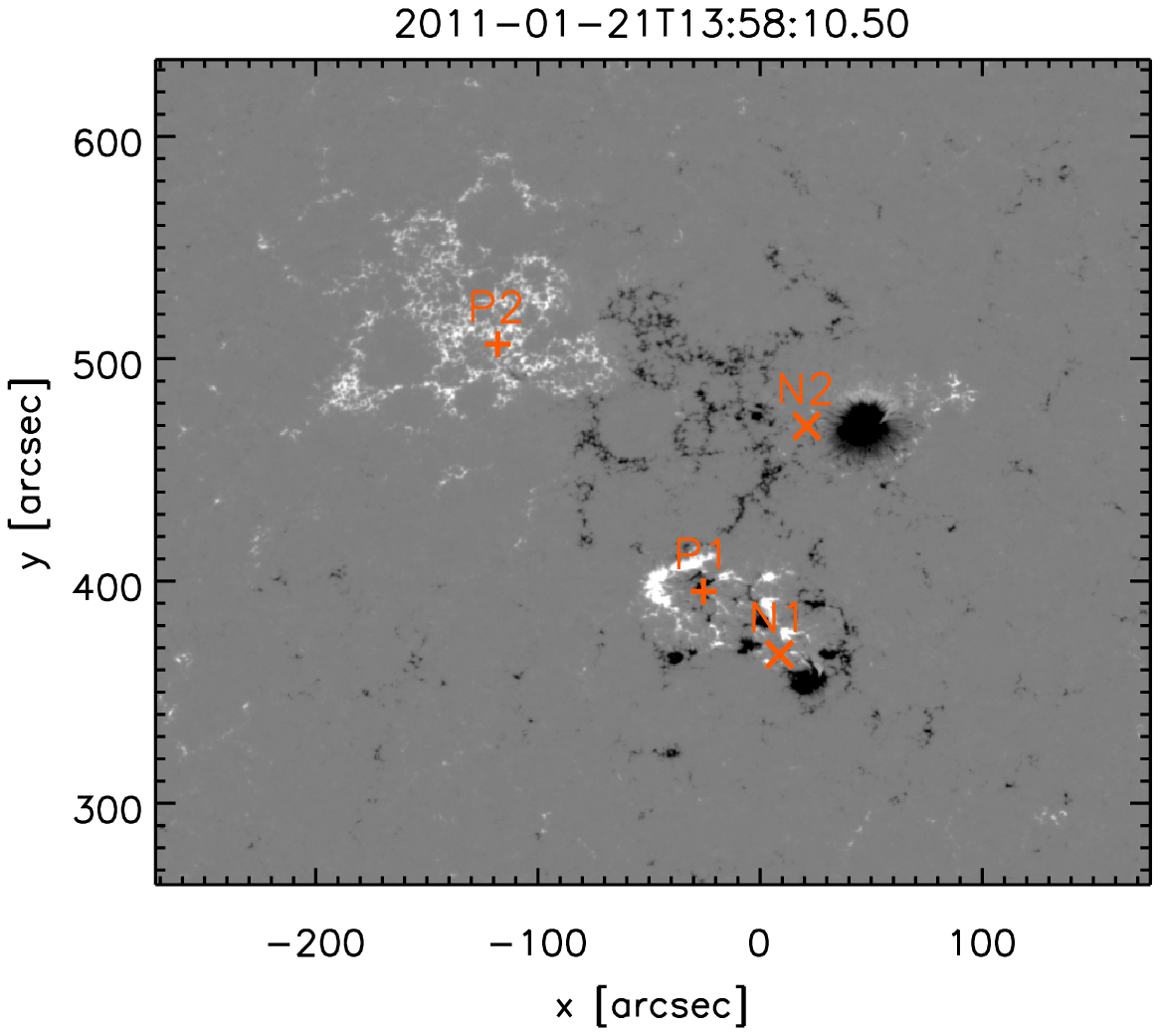}
\plotone{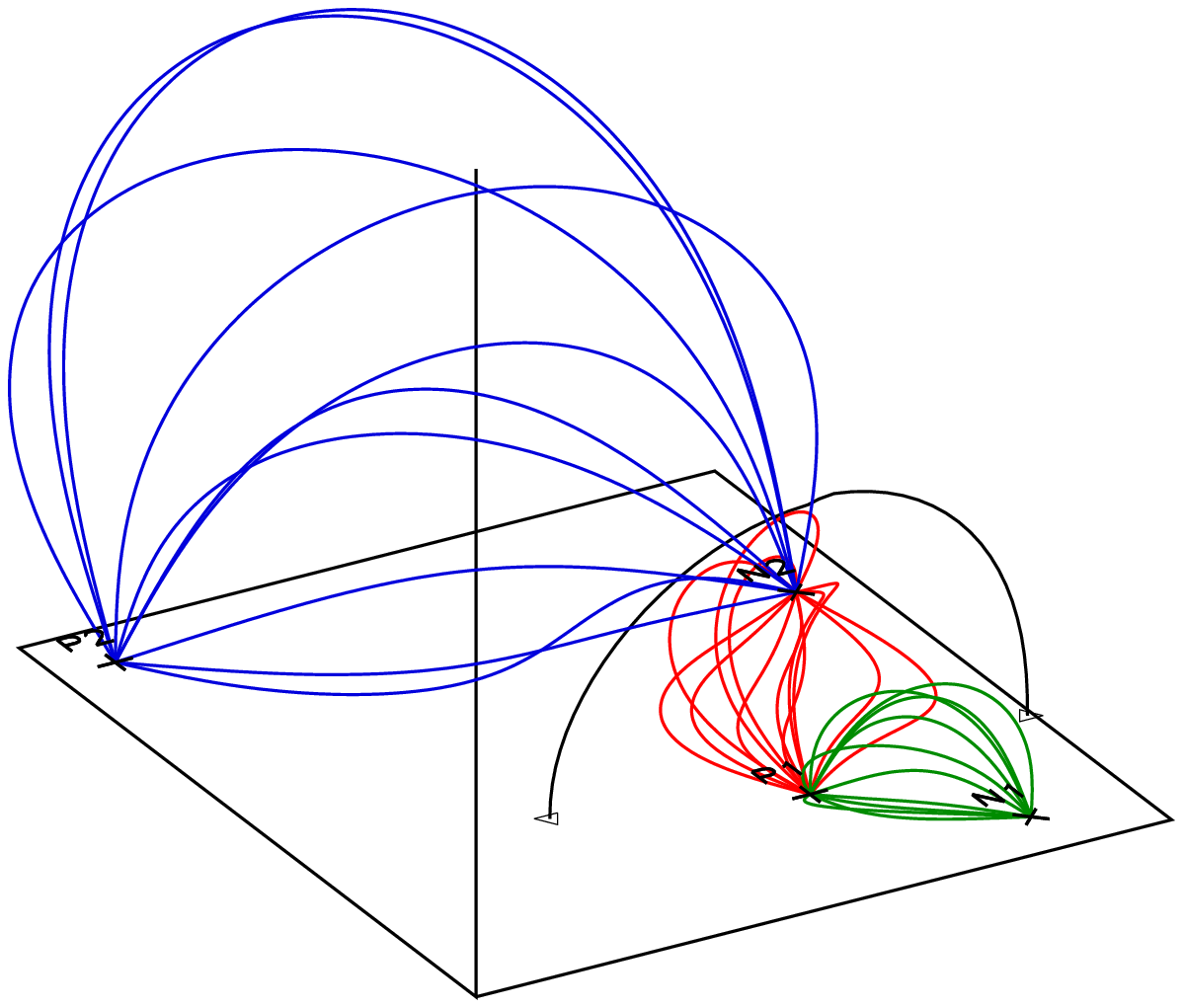}
    \caption{Top: Using the line-of-sight magnetogram from SDO/HMI, we build a MCT model (orange).  The magnetogram is shown on a linear scale that saturates at $\pm700$ G.  Bottom: Three out of four total magnetic domain are illustrated: the existing AR (blue), the emerging AR (green), and the interconnecting flux between the new and old ARs (red).  The separator, the junction at which the four magnetic domains meet, is shown in black.}
    \label{fig:qp}
  \end{figure}
  
We construct a model of the magnetic field of the emerging and existing active region system in order to precisely define the interconnecting flux between these two magnetic systems.  Because it is common to assume that an active region is a distinct, internally connected magnetic system that emerges from within the Sun \citep{1994ApJ...436..907F,2000SoPh..192..119F}, any flux linkage between two ARs must be formed by some process post-emergence.  We construct a Magnetic Charge Topology (MCT) model which uses magnetic ``charges'' to represent the AR's polarities \citep{1996SoPh..169...91L}.  A potential field is subsequently extrapolated from these to create a volume-filling field model.  To do this, we utilized the MPOLE suite written in IDL \citep{1996SoPh..169...91L}.  As a potential field is the state of lowest possible magnetic energy, this simple case provides a good basis for comparison with the EUV imaging observations.

The magnetic charges in our model are constructed from a series of line-of-sight magnetograms from the  {\tt hmi.M\_720s} data series of SDO/HMI \citep{Scherrer2012}.  A subregion which contains both active regions is selected from the full-disk magnetograms.  Flux concentrations are automatically identified from a smoothed version of the extracted magnetogram; the filtered field is the vertical field from a potential field extrapolation to a height of 6 pixels above the magnetogram surface \citep{1996SoPh..169...91L}.  Local extrema are identified and grouped with all surrounding, downhill pixels whose magnitudes are greater than 75 G.  Regions smaller than 200px in total pixel count are then discarded, to better isolate the regions of strong flux from the small flux concentrations that are scattered all over. Each flux concentration is then replaced by a magnetic charge located at its flux-weighted centroid, with net charge given by the integrated flux.  This step also includes introducing a factor that accounts for the projection of the line-of-sight field being a component of the radial.  The results of this stage are referred to as the ``many-poles'' model.

We choose to reduce our many-poles model to a quadrupole in order to clarify and simplify the emergence scenario.  As the flux concentrations are continuously combining and separating, there is no guarantee that any one pole exists continuously in the many-poles model.  \cite{Tarr2012} took steps to remedy this by utilizing automated tracking algorithms to gain consistency as the sources evolve in time.  Here, at the expense of losing some of the finer topological detail about the two-AR system, we gain continuity by reduction to a quadrupole.  Loss of resolution is not a major problem since we are concerned with interconnecting flux between the two regions.  This information is retained in the simplification, as the emerging and existing AR polarities are still distinct. 

To reduce the many-poles model to a quadrupole, we manually trace a boundary around the emerging active region at each time in the sequence.   All poles within the boundary are reduced to a positive-negative pair, P1 and N1, by totaling the charges of each sign in the region and computing the flux-weighted centroid of them to determine the single charge's location.  The poles outside this boundary are reduced similarly to P2 and N2.  
After the quadrupole reduction, the field is organized into four magnetic domains: P1-N1 (emerging), P2-N2 (existing), P1-N2 (interconnecting), and P2-N1 (also interconnecting).   The global field is extrapolated from the point sources at the photosphere and field lines are traced from a positive source to its termination at a negative source.  As the quadrupole is not exactly flux-balanced, this can include sources at infinity.  However, as we are concerned only with the AR-AR interaction, we consider the four topological domains previously mentioned for clarity. 
The first three domains can be seen in Figure \ref{fig:qp} (bottom), along with the separator (black) which is the shared boundary between the four domains. For clarity, we do not show the fourth domain.

\begin{figure*}[]
\centering
\plotone{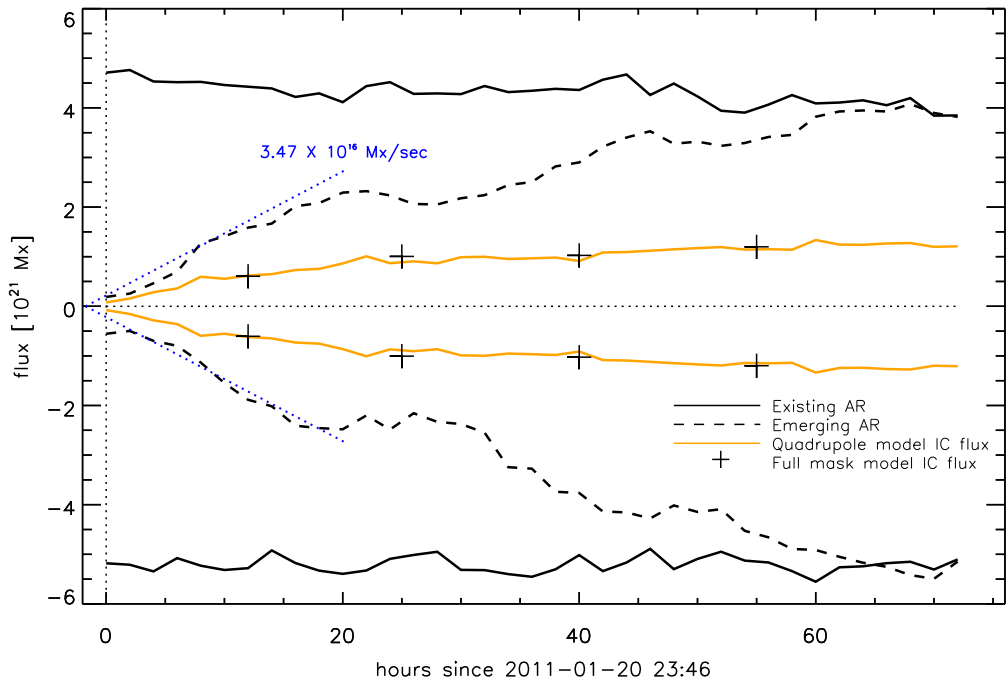}
    \caption{The fluxes from our MCT model are shown here.  The flux of the active regions are shown in black: the flux of existing AR (solid) is fairly constant in time while the emerging AR flux (dashed) increases over a ~72 hour period.  The interconnecting flux from the quadrupolar MCT model is shown in orange.  The crosses are the interconnecting flux determined from many magnetic charges before the reduction to a quadrupole to make sure we did not lose information by doing the reduction of the model.  The vertical dotted line denotes the first time of the MCT modeling and the x-intercept of the blue dotted linear fit denotes what we call the time of emergence of the new active region.}
    \label{fig:potentialflux}
\end{figure*}

Creating a quadrupolar model of the active regions' polarities over a time period captures the emergence of the P1-N1 system and provides flux measurements which are used to characterize the active region emergence in this event.  From mid-emergence (2011-01-21T09:58:10), we work backwards in time until the polarities from the emerging active region do not fit the criteria for the flux concentration selection (2011-01-20T23:46:10).  This provides one crude definition of the onset of emergence.  This model-building was done every 2 hours for 72 hours, until the flux of the emerging active region began to level-off.  During this time period, the P2 and N2 poles stay fairly constant in both their individual magnitudes and their differences -- they are not quite flux balanced, but differ by a median value of $8.9 \times 10^{20}$ Mx (about 10\% of the flux in either polarity).  To determine the beginning of our emergence event, the rate of flux emergence in the P1-N1 active region was estimated by eye such that it reasonably fit the rate of both the positive and negative polarity emergences.  We find that these fluxes grow at approximately $3.47 \times 10^{16}$ Mx s$^{-1}$ during the initial phase (from 2011-01-20T23:46 to 2011-01-21T19:58:10.50, Figure \ref{fig:potentialflux}, blue dotted line).  We fit a straight line to the rate of change of the unsigned flux in each polarity; this line intersects the horizontal axis in Figure \ref{fig:potentialflux} on 2011-01-20 at 22:01.  We will henceforth call this time the beginning of the emergence.

From this quadrupolar model, we calculate the interconnecting (IC) flux that connects the emerging positive (P1) to the existing negative (N2) polarity.  
This is the same as the flux through the surface bounded by the separator -- the intersection of the separatrices, surfaces which divide topologically distinct regions -- provided the separator curve is closed such that the P1 and N2 charges lie on opposite sides of this bounded surface \citep{2004ApJ...608.1106L}.  We perform this calculation using a Monte Carlo method that uses randomized locations around each source to trace field lines \citep{2005ApJ...629..561B}.  The ratio of number of field lines that connect charges P1-N2 to the total amount of field lines traced around these charges determines the fraction of flux that connects them and thereby determines the P1-N2 interconnecting flux.  In Figure \ref{fig:qp} (right), the red field lines belong to the interconnecting domain. Observe that these field lines go through the surface bounded by the separator (black).  Figure \ref{fig:potentialflux} shows the results for the quadrupolar modeling over the emergence time in orange.  The crosses shown in Figure \ref{fig:potentialflux} are instances where the interconnecting flux was determined with the many-poles model.  When using the more complex model, the more complex topology requires us to consider the contributions from multiple poles between active regions which is much more computationally intensive.  With the quadrupolar model, we only need to consider the P1 and N2 poles.  Both the many-poles and quadrupolar models give approximately the same result and thus we may use the computationally cheaper method at other times as well.

The interconnecting flux changes in time throughout our model.  A time-evolving flux is described by Faraday's law
\begin{equation}
\dot{\Phi} = - \oint \vec{E} \cdot d\vec{l} = -\mathcal{E}\label{eq:faradays}.
\end{equation}
The presence of an electric field is of no consequence in a vacuum, since any electric field is allowed to be aligned with the magnetic field \citep{1998SoPh..179..349L}.  In a conducting plasma, however, this is not permitted and the electric field is required to be perpendicular to the magnetic field \citep{1998SoPh..179..349L}.  Thus there must be a difference in the mechanism causing the evolving interconnecting flux between our idealized, MCT model in vacuum and reality in a conducting plasma.  This mechanism is magnetic reconnection.  A perfect conductor would not break magnetic field lines to rearrange the field topologically; the presence of the EMF, $\mathcal{E}$, violates the condition of a good conductor.  Therefore, reality lies somewhere in between zero interconnecting flux and the interconnecting flux obtained through our MCT model.  In the next section, we seek evidence of newly reconnected flux between the old and new active regions in coronal loops using data from SDO/AIA to make measurements of this reconnection process.

\section{Interconnecting Flux Evident in AIA}\label{sec:data}

  \begin{figure}[]
\centering
\plotone{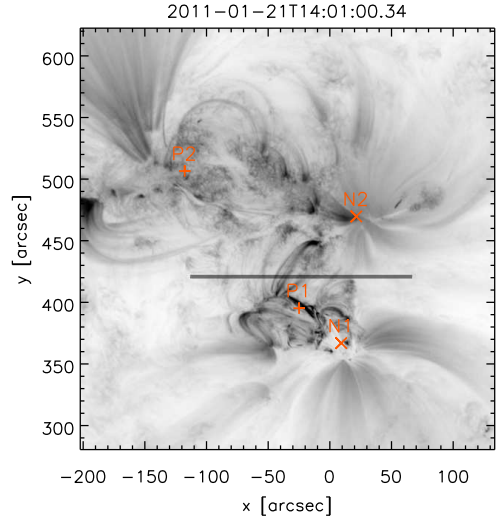}
    \caption{A $171$\AA\ image from SDO/AIA showing the emerging and existing active regions.  The slit location is shown by the dark bar.  We assume that any coronal loops connecting the two ARs will cross that slit.  The figure is displayed with a reverse (light-to-dark) color scale, has been square root-scaled as to better see the coronal loops for reference, and saturates at $55 (\mbox{DN/s})^{-1/2}$.  The quadrupolar MCT model is overlaid for reference in orange.}
    \label{fig:slitqpover}
  \end{figure}

We use a series of SDO/AIA images to measure magnetic reconnection between the emerging and existing active regions.  In most theories reconnection is accompanied by an energy release \citep[e.g.][]{1958NCim....8S.188S,PhysRev.107.830,1972ApJ...174..499P}.  We expect the plasma that traces the coronal magnetic field to be heated as a result of this reconnection and become bright in the EUV.  We go further and assume that the only means by which loops are heated, and made visible, is the energy released by the reconnection which creates interconnecting flux.  This allows us to correlate the reconnected magnetic flux with the magnetic flux in newly brightened interconnecting loops.  \citet{2005ApJ...630..596L} used this same approach to quantify separator reconnection between an emerging AR and the overlying field in the form of coronal loops between the two ARs.  Magnetic reconnection is occurring in the present emergence event as well:  we see the coronal loops between the two active regions in the AIA $171$\AA\ images (see Figure \ref{fig:slitqpover}).  This is clear evidence of magnetic reconnection in the corona, and with the foregoing assumption we can quantify its rate.

We make measurements of magnetic reconnection by quantifying the accumulated area of coronal loops crossing the surface bound by the separator at a given time.  Ultimately, we obtain accumulated flux, manifesting itself in newly appeared interconnecting loops.  By measuring the rate of change of the interconnecting flux, we determine the integrated reconnection electric field using Faraday's law (Equation \ref{eq:faradays}).  To make this measurement we use 48 hours of AIA data in the 171\AA{}  channel with a 1 minute cadence, starting at the time we identified in \secref{sec:MCT} as the beginning of the emergence.  Although evolution of the emerging flux region continues beyond this, 48 hours was enough time to assure that measurable reconnection had taken place.  The data were processed with the SolarSoft routine {\tt aia\_prep} to convert level 1 data to level 1.5 \citep{FreelandHandy1998}.  This included exposure normalization to convert intensities from DN to DN/sec.

\subsection{Cataloging Coronal Loops}\label{sec:catalog}

\begin{figure*}[!p]
\centering
\includegraphics[width=0.7\linewidth]{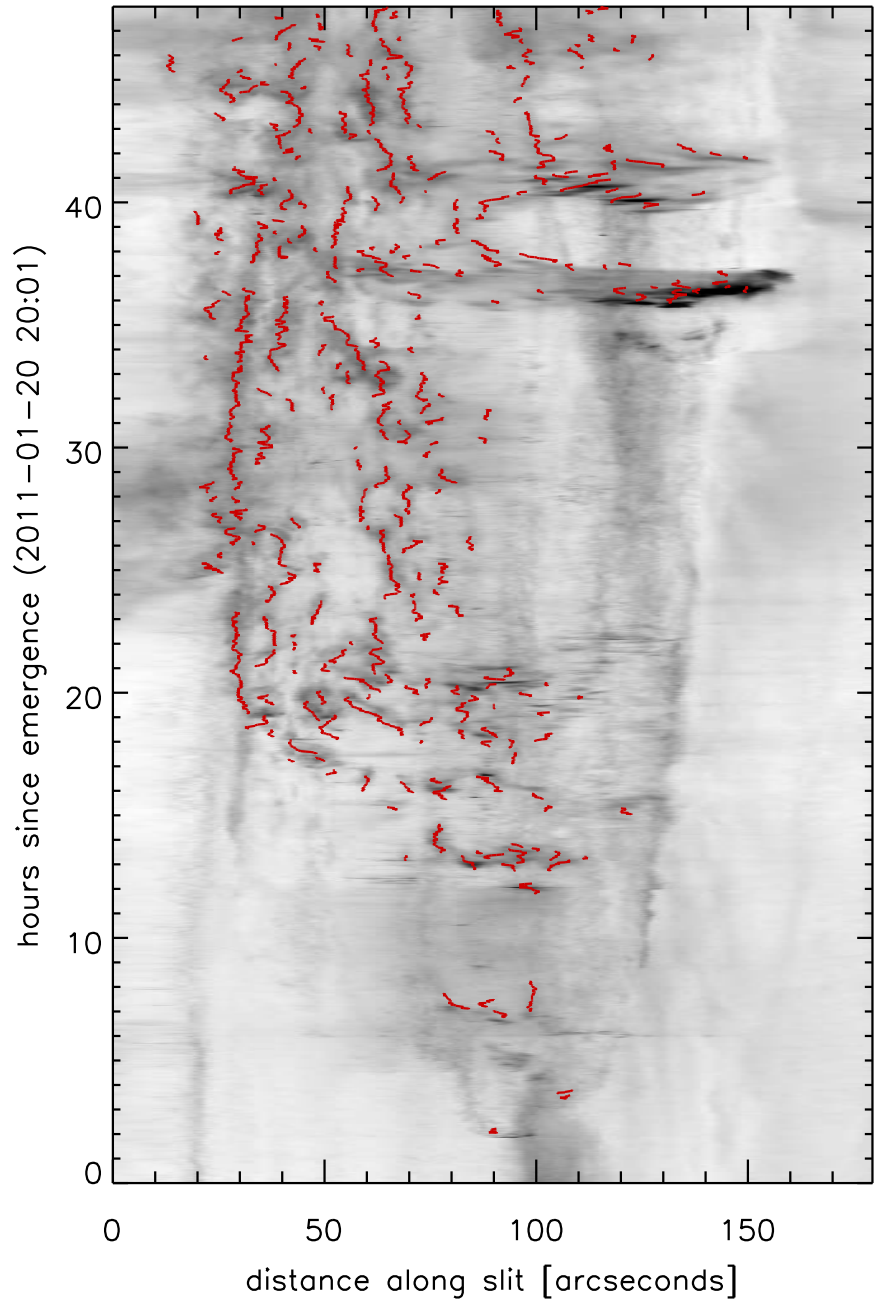}
\caption{A time/space stack plot built from the virtual slits extracted from AIA images.  Verified interconnecting coronal loops are plotted over the stack plot in red.  The stack plot has been square root-scaled for this figure to better distinguish the streaks of strong emission; the data used in the analysis was not scaled in this way.  The image's reverse color scale saturates at $55 (\mbox{DN/s})^{-1/2}$.}
    \label{fig:stack}
\end{figure*}

\begin{figure}[]
\centering
    \plotone{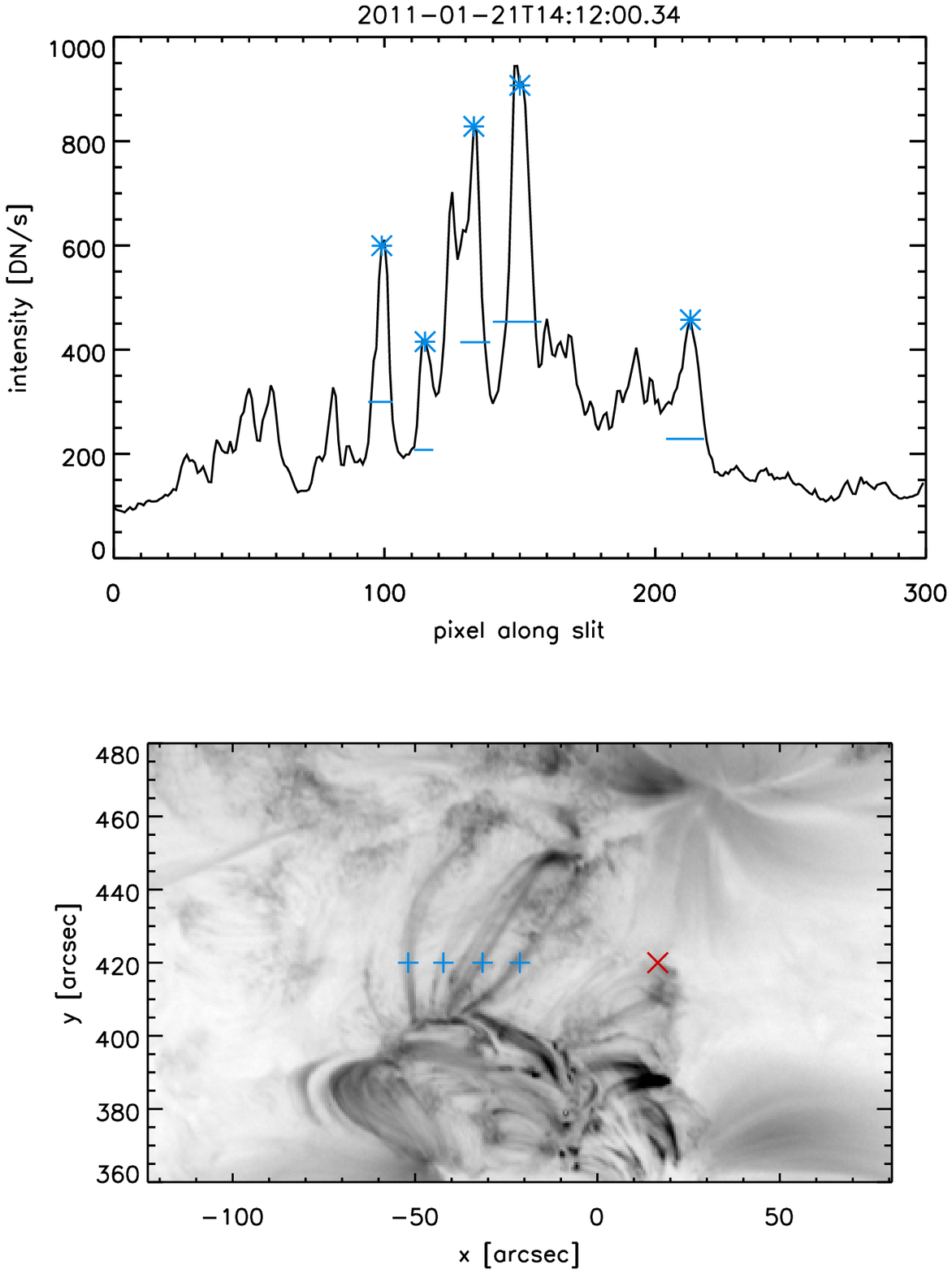}
    \caption{Top: A single row from our stack plot.  Peaks from our local maxima finder are shown as asterisks and the diameters are shown as horizonal bars.  \\ Bottom:  The peaks identified from the top panel are now plotted over the AIA image.  The peaks indicated by blue crosses do appear to correspond to interconnecting coronal loops.  The red X does not, and would be discarded.  The reference AIA image has been square root-scaled for better clarity in seeing the coronal loops in the data.  The reverse color scale of the image saturates at $60 (\mbox{DN/s})^{-1/2}$.}
    \label{fig:bumps}
\end{figure}

A time/space stack plot is used in our analysis to catalog all the coronal loops that connect the two ARs.  An array of pixels was extracted from each of the prepped AIA images (Figure \ref{fig:slitqpover}).  This was a horizontal array 300 pixels (180'') wide and 5 pixels tall.  The slit is positioned between the two ARs so that any interconnecting loop would be certain to cross it and no loops internal to ARs would.  That array is averaged to a 300x1 pixels array, where each pixel was the mean of the 5 in the column extracted from the data.  At the beginning of the data set (2011 January 20 at 22:01 as determined from the linear fit in \secref{sec:MCT}), the lower left pixel of the slit was located at [-250'',420''] in Cartesian coordinates on the plane of the sky, where disk center is [0'',0''].  The x-position was tracked with solar rotation thereafter in the data set.  The intensity arrays were stacked to build the stack plot shown in Figure \ref{fig:stack}. As the coronal loops are bright, they will appear as local maxima in intensity in one row of the stack array, as seen in Figure \ref{fig:bumps} and will appear in the stack plot as persistent bright streaks in time\footnote{Figures \ref{fig:slitqpover}, \ref{fig:stack}, and \ref{fig:bumps} use a reverse (light-to-dark) color scale and thus our ``bright streaks'' of emission are dark in this color table.  We choose to refer to the appearance of these loops from the onset of intense emission as ``brightening'' for consistency with common usage}.  

Bright streaks are identified from the stack plot automatically, and subsequently visually confirmed to be coronal loops.  For each time slice of the stack plot, bright peaks along the slit are determined.  The intensity is boxcar smoothed with a kernel of 5 pixels along the artificial slit.  A threshold is determined from taking an average of the intensity values along the slit; all pixels below the threshold are removed from consideration.  Local maxima are then identified and the loop's ``edges'' are determined from their full width at half maximum (FWHM).  From the identified edges, we obtain the diameter.  A peak must be greater in smoothed intensity than 90 DN/s above the threshold to be included in the set of local maxima.

Successive intensity peaks at a similar location in the slit are linked together to turn individual peaks into sustained bright streaks, which are candidates for being interconnecting loops.  In order to be linked, the subsequent peak (1 minute later) can have moved no more than 3 pix (1.8'') to either side.  This is within a typical loop radius of 2'' in previous studies with TRACE \citep{2000ApJ...541.1059A,2005ApJ...630..596L} and AIA 
\citep{2013SoPh..283....5A}.  When there is no next peak to be linked to, this is the end of the bright streak.  If in a given row there is a new maximum that does not link to any existing, ``alive'' loops, then that is considered the beginning of a new bright streak.  We allow for a peak to temporarily disappear for one time step only, such that two bright streaks at the same location and which look to be the same loop but are separated by one slice in the stack plot are not counted as two distinct loops.  There must be two successive images without linked peaks for a bright streak to truly end.

The properties of the streaks are interpreted as properties of loops.  Through our maxima-linking procedure, we obtained both a lifetime (taken from the start and end of the bright streak in the stack plot) and a median diameter (median of all FWHMs during maxima identification) for every bright streak.  At this point in the automatic selection process, we have 2259 loop candidates.  Note that these candidates included 852 maxima that were not linked to another at a different time, and 394 streaks that only lasted 2 time steps.

Of the 2259 bright streaks automatically selected, only 554 lasted 5 minutes or longer.  This minimum-lifetime cutoff of 5 minutes was chosen to make the subsequent manual-confirmation from a data set that did not include ephemeral bright streaks.  We visually examine each of the remaining loop candidates to ensure that it is a true coronal loop.  The automatic maxima finding and linking procedure identifies many candidates and these can include features other than coronal loops, for example transition region moss \citep[see][]{1999SoPh..190..409B}, that produce persistent bright streaks.  For example, the right-most peak in the top panel of Figure \ref{fig:bumps} is caused by the chromospheric moss feature marked with a red X in the lower panel.  We use this kind of visual inspection to identify and discard bright streaks that are not loops.  After this process, we have 301 verified interconnecting coronal loops, which are shown on the stack plot in Figure \ref{fig:stack} in red.

\subsection{Loop Properties}\label{sec:loopprop}
\begin{figure}[]
\centering
    \plotone{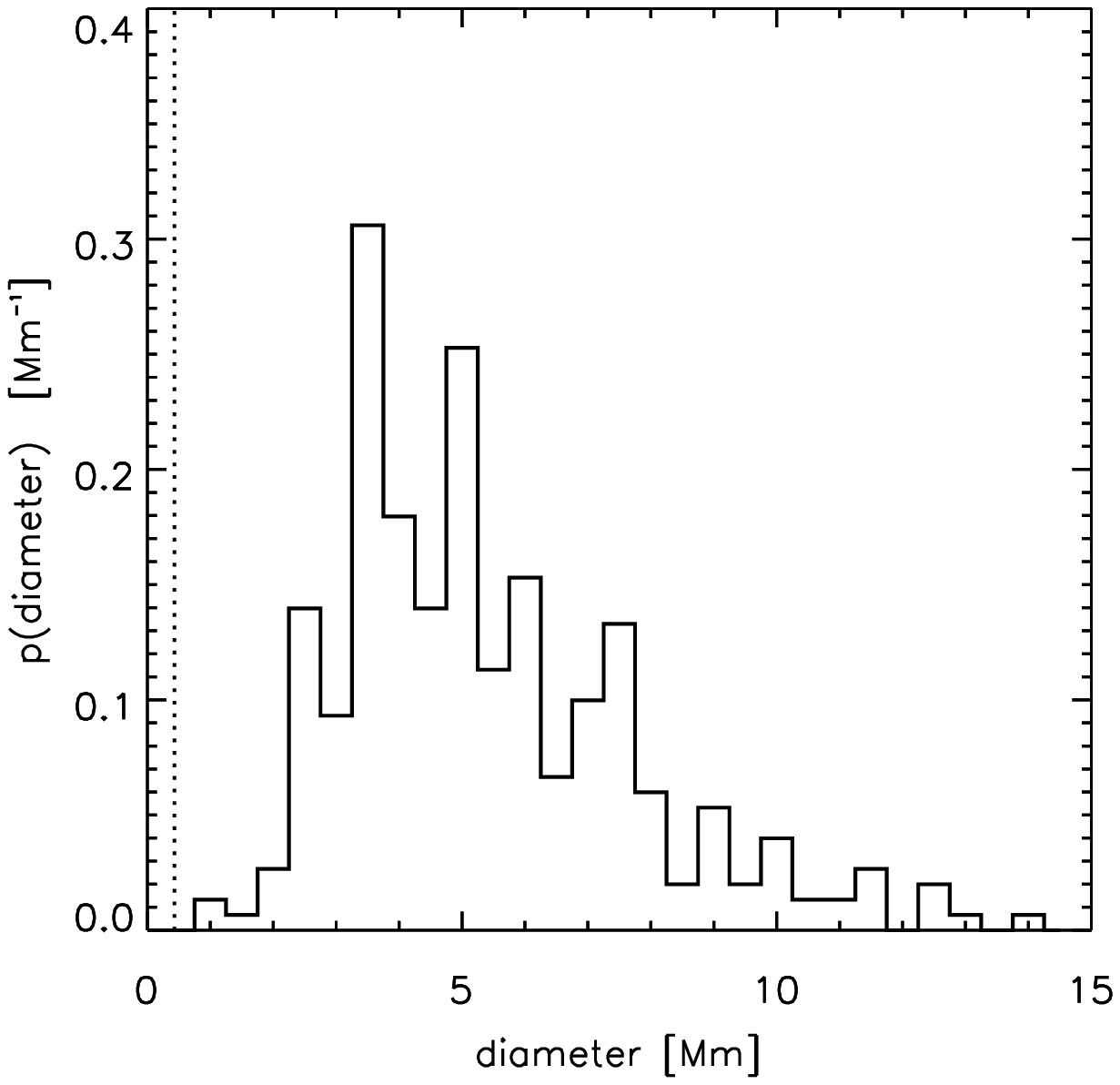}
    \plotone{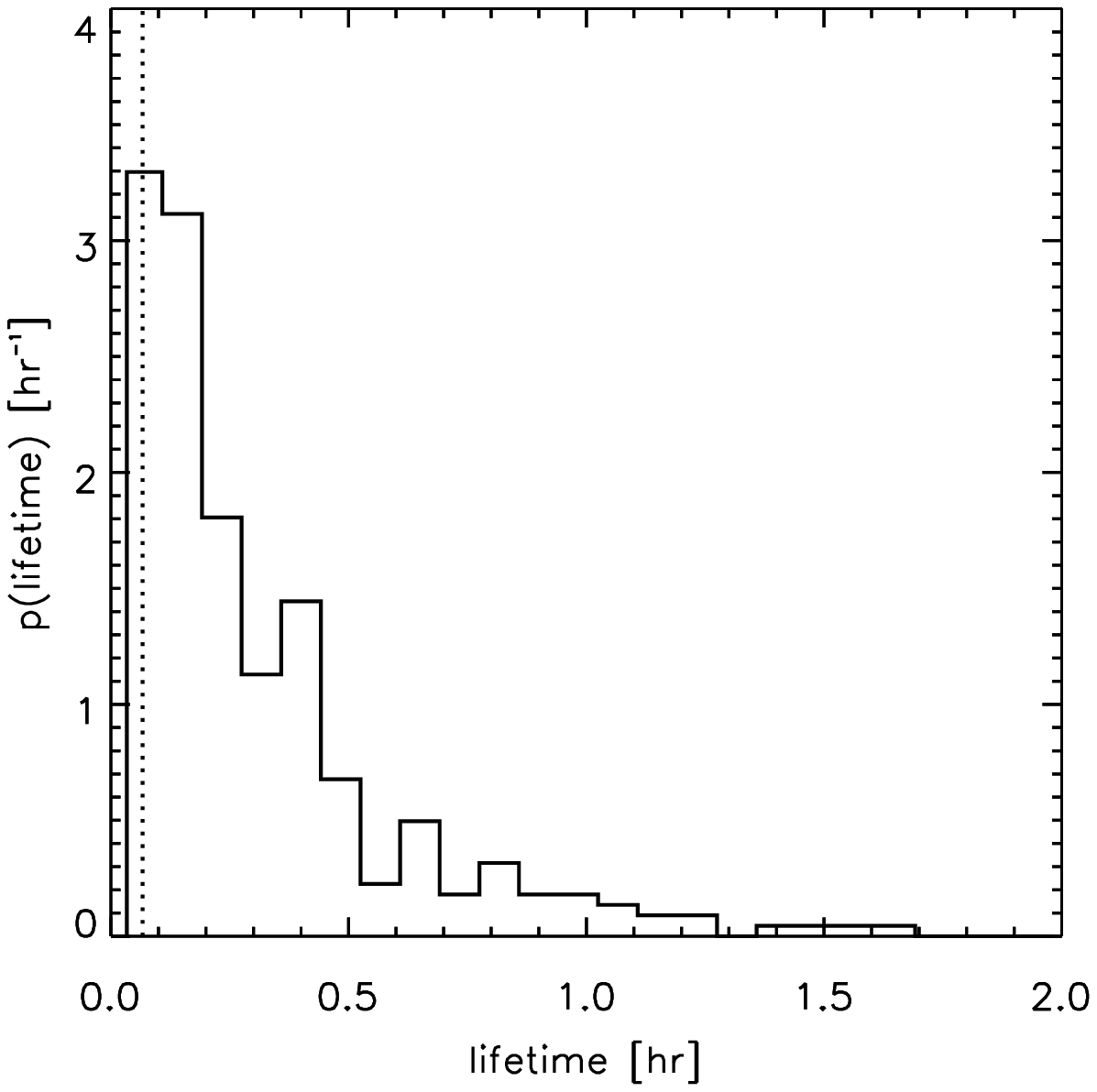}
    \caption{Histograms denoting the properties of the 205 loops interconnecting the two ARs.  Top: Median loop diameters in Mm.  The dotted vertical line denotes the size of one AIA CCD pixel during this event (0.6'', 0.43 Mm).  Bottom: Loop lifetimes in minutes.  This histogram cuts off one data point to better show the bulk of the distribution for clarity, whose lifetime of 202 minutes was much greater than the rest of the distribution.  A vertical dashed line shows the threshold time (5 min) for a loop to be included in our database.  The abrupt drop at the lower end of the lifetime distribution is an artifact of our loop cutoff criterion.}
    \label{fig:lifediam}
\end{figure}

Each of the 301 identified loops is characterized by properties obtained from the stack plot.  Lifetimes are determined from the stack plot slices which contain the birth and death of the loop.  This is the length of time that there is a continuous (in time) peak in the intensity at nearby locations along the virtual slit.  A loop's diameter is found from the median value over all times of the FWHM found in the local maxima-finding process.  The histograms in Figure \ref{fig:lifediam} show these values\footnote{The lifetimes histogram cuts off a single outlier data point, a loop that lasts 202 minutes}.  The shoulder on the low end of the lifetime distribution is an artifact of our minimum lifetime cutoff criteria.  Loops have a median diameter $5.9$ Mm (median of all medians), and the loops of the smallest diameter are well above the CCD pixel size of AIA (approx. 0.4 Mm).  Loops usually last less than $30$ minutes with a median lifetime of $15$ minutes.

\begin{figure*}[]
\centering
    \plotone{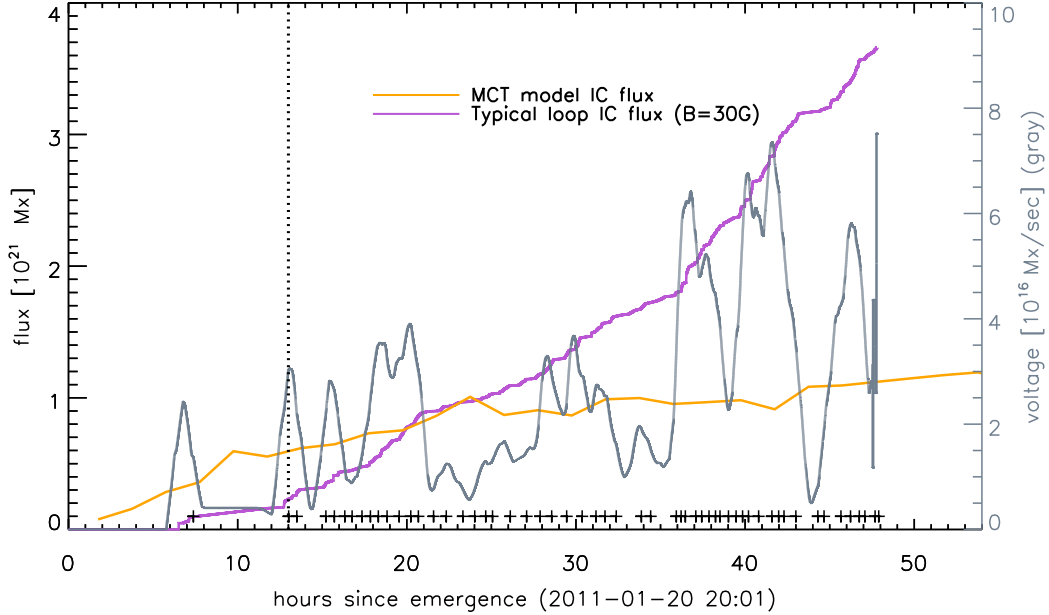}
    \caption{Integrated reconnection electric field.  The mean voltage is $2.13 \times 10^{16}$ Mx/sec $= 213$ MV.  Crosses at the bottom denote the times used for magnetic modeling in \secref{sec:annymodel}.  The vertical dotted line denotes when the bulk of reconnection begins, a 13 hour delay from the emergence.}
    \label{fig:fluxvoltage}
\end{figure*}

We compare these observed lifetimes to a typical cooling time of a coronal loop.  During its cooling phase, the loop will cool through a combination of radiation and thermal conduction.  These processes operate on time scales given by 
\begin{eqnarray}
\tau_{\rm rad} ~=~ 2.3\times 10^3\,{\rm s}\, \left( {T\over 10^6 \,{\rm K} }\right)^{3/2}
\left( {n_e\over 10^9 \,{\rm cm}^{-3} }\right)^{-1} \\ \nonumber
\tau_{\rm cond} ~=~ 270 \,{\rm s}\, \left( {n_e\over 10^9 \,{\rm cm}^{-3} }\right) \left( {T\over 10^6 \,{\rm K} }\right)^{-5/2} \left(L \over 10^9 \,{\rm cm}\,\right)^{2} 
\end{eqnarray}
where $T$ and $n_e$ are the temperature and electron density at the tube's apex, and $L$ is its full length \citep{1995ApJ...439.1034C}.  When the loop is sustained by uniform volumetric heating \citep[i.e. RTV equilibrium,][]{1978ApJ...220..643R} or is cooling quasi-statically, conductive and radiative cooling rates are equal.  This condition can be used to eliminate the electron density, $n_e$, and obtain a combined cooling time
\begin{equation}
\frac{1}{\tau} = 
\frac{1}{\tau_{rad}} + \frac{1}{\tau_{con}} = 
\frac{2}{\tau_{rad}} = 
\frac{2}{\sqrt{270 \cdot 2.3\times10^{3}} T_{6}^{-1/2} L_9},
\end{equation}
giving a cooling time 
\begin{equation}
\tau = 6.5 \, L_9 T_6^{-1/2} \,{\rm [min]},
\end{equation}
where $L_9=L/(10^9\, {\rm cm})$, and $T_6=T/(10^6\, {\rm K})$.
For a loop of length $60$ Mm ($L_9 = 6.0$) and observed in the 171$\AA\ $ passband at $T=0.6$ MK ($T_6 = 0.6$), the cooling time $\tau = 50.3$ minutes.  The lifetime range for our identified loops are typically of the same order of magnitude as this crude estimate.  This is consistent with loops which have been heated impulsively, for example by reconnection, and then undergo free cooling.

Our 301 loops have properties similar, but not identical, to the 43 found by \cite{2005ApJ...630..596L}, also during a 48-hour interval.  The typical loop diameter we find ($5.9$ Mm) is approximately $60\%$ larger than the value from \cite{2005ApJ...630..596L}, $3.7$ Mm.  It is possible the discrepancy arises from the artificial enhancement of our diameters due to the boxcar smoothing.  The lifetimes found between that work and this one lie mostly within 90 minutes (1.5 hours), with both studies containing several loops that last much longer.  As in \cite{2005ApJ...630..596L}, there was a delay between emergence and the bulk of reconnection.  The delay in this case was 13 hours since the beginning of the emergence as determined in \secref{sec:catalog} (the vertical dotted line in Figure \ref{fig:fluxvoltage}), compared to 24 hours in \cite{2005ApJ...630..596L}.  Delays in the range of 12 to 25 hours have previously been observed \citep{2005ApJ...630..596L,2008A&A...488.1117Z,KobelskiThesis,Tarr2014} but more investigation is required to confirm whether this seemingly common phenomena is ubiquitous in flux emergence.  If we find that it is, we also wish to ascertain what properties of the system determine if the delay is closer to 12 hours, as in this work, or to the 24 hour interval found in \cite{2005ApJ...630..596L}.

We use the loops and their properties to calculate the time-varying interconnecting flux.  Under the assumption that the flux tube is a cylinder, we turn the identified diameters into cross-sectional areas.  We define a cumulative area, $A(t)$, which jumps by the cross-sectional area where it crosses the slit of a particular loop at the instant that loop is born.  To obtain interconnecting flux as a function of time, we multiply $A(t)$ by a characteristic magnetic field strength $B_{char}$.  We chose $B_{char} = 30$ G as informed by our MCT model.  We use the quadrupole model to find a typical field strength around the approximate location of the separator in the middle of our data set; reconnection occurs along that boundary to exchange flux between the magnetic domains.  Then, interconnecting flux as a function of time is 
\begin{equation}
\Phi(t) = \int \mathbf{B_{char}} \cdot d\mathbf{A(t)} = B_{\text{char}} A(t),
\label{eq:flux}
\end{equation}
which is shown as the purple line in Figure \ref{fig:fluxvoltage}.

The integrated reconnection electric field was determined from Faraday's law (Equation \ref{eq:faradays}) by taking a time derivative with respect to $\Phi(t)$.  The voltage, $\dot{\Phi}$, is shown in Figure \ref{fig:fluxvoltage} as the gray line.  Though this has been boxcar-smoothed over half-hour time intervals to eliminate some of the high frequency variation for clarity, this reconnection between the old and new ARs occurs in bursts similar to the observations of \cite{2005ApJ...630..596L}, impulsively heating these loops into the 171\AA\ passband.

The interconnecting flux obtained from accumulated coronal loops is larger than that from the MCT model presented in \secref{sec:MCT}.  This is despite possible under-sampling of the coronal loops between the two active regions during our cataloging process by choosing a minimum loop lifetime of 5 minutes, as well as the fact that we are using 171\AA\ data only.  While the effect of reducing our system of two active regions to a quadrupole was minimal with regards to the calculation of interconnecting flux, a more complete model of the field might lead to a characteristic field different than we obtained through our MCT model earlier in this section.

\section{Non-potential force-free field modeling}\label{sec:annymodel}

\begin{figure}[]
\centering
\plotone{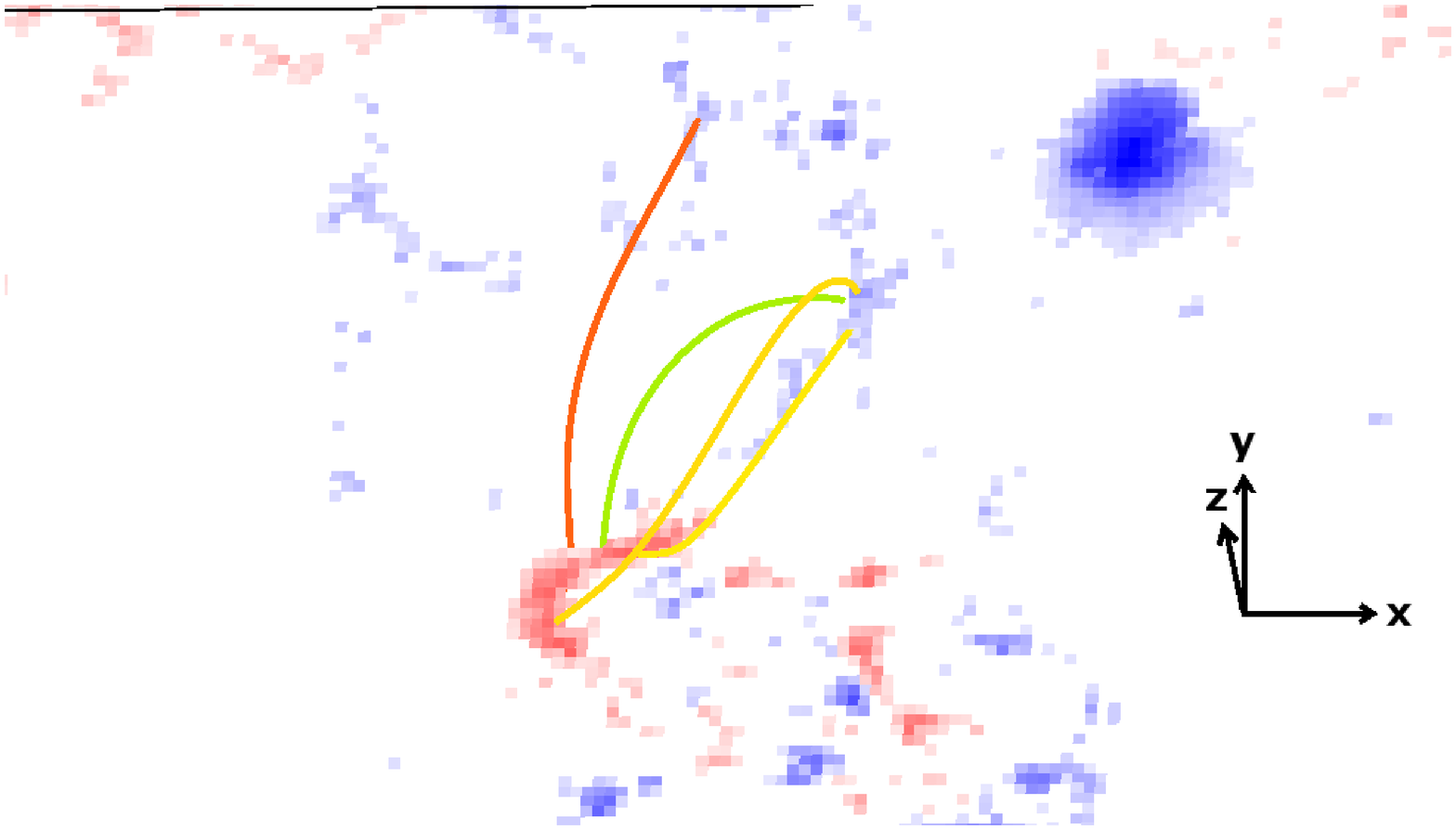}
\plotone{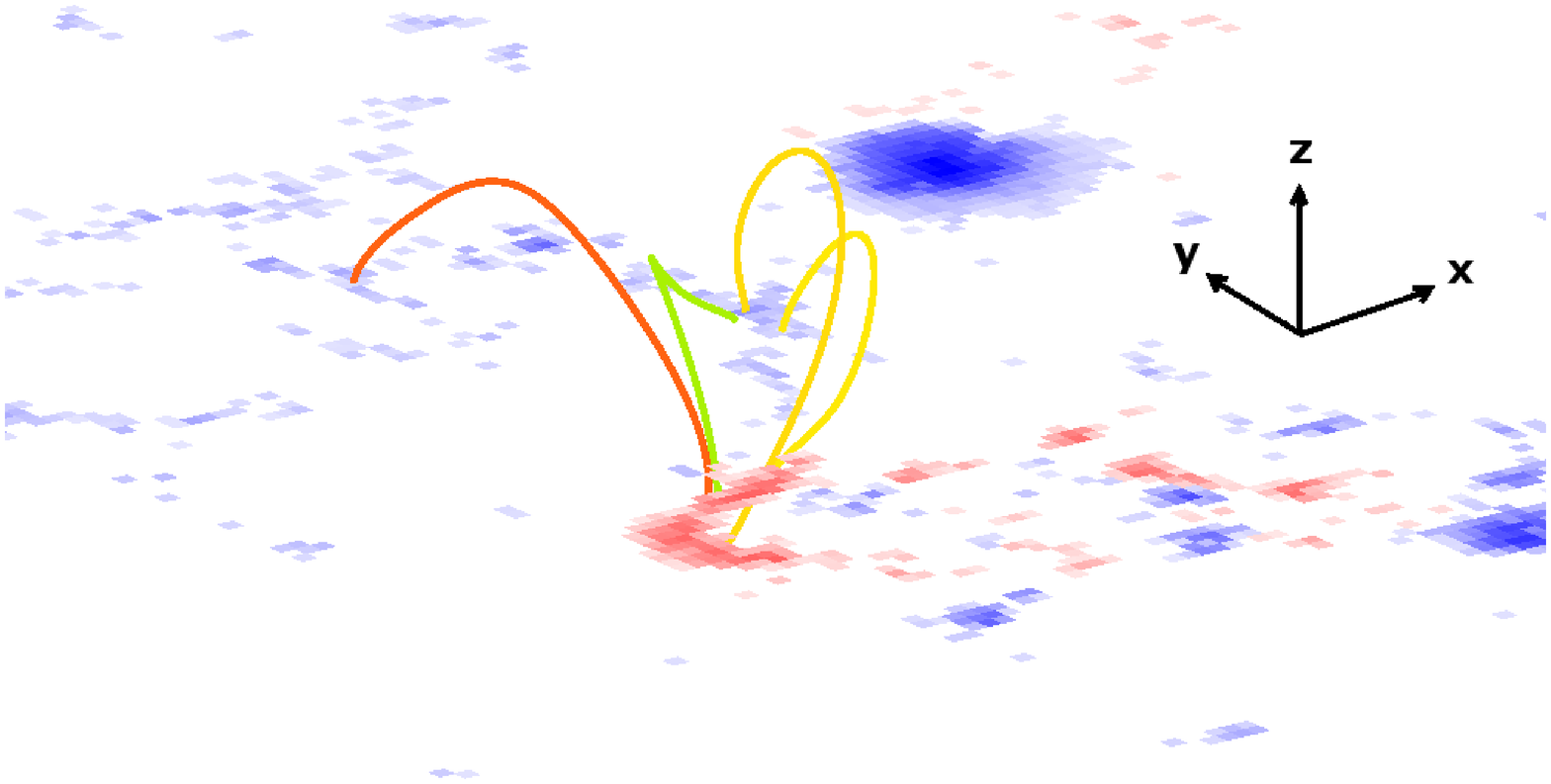}
\plotone{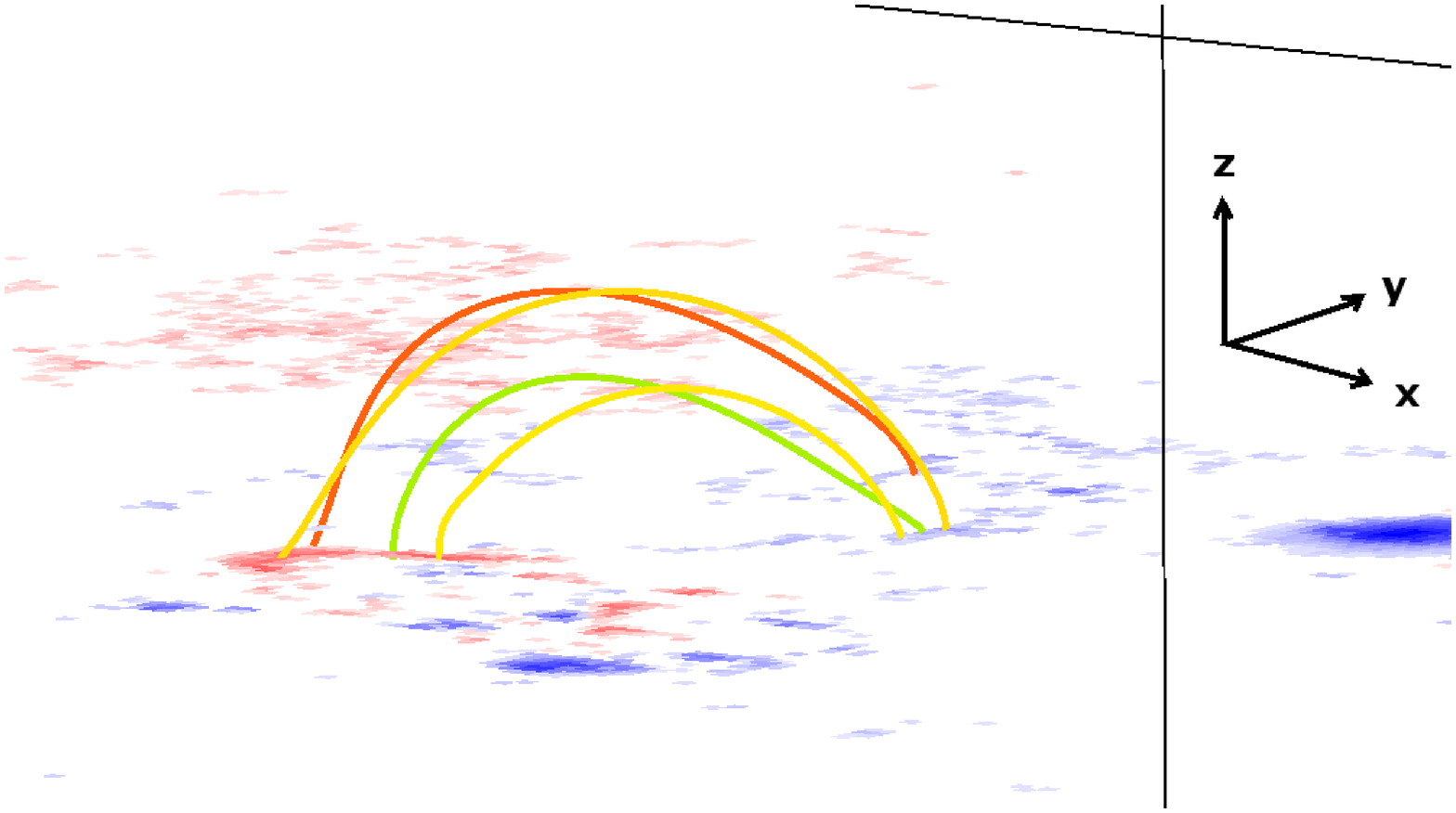}
    \caption{A 3D rendering of the loops at 2011-01-21T14:21 (compare to Figure \ref{fig:bumps}) that were reconstructed from our LFFF models.  The top figure shows a line-of-sight viewing angle, similar to how the loops are seen in AIA.  The lower panels show the three-dimensional reconstruction of the loops at different angles.}
    \label{fig:3dmodel}
\end{figure}

\begin{figure}[]
\centering
\plotone{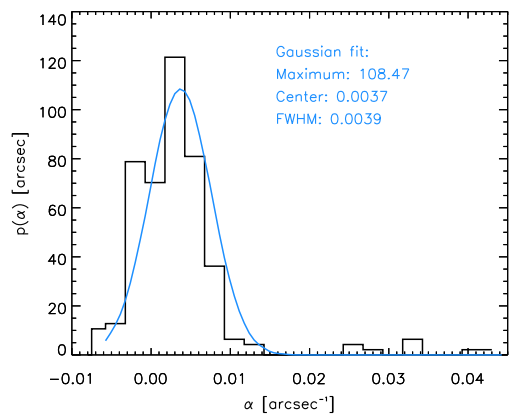}
    \caption{A histogram showing the $\alpha$ values for all the loops obtained through our LFFF model.  The Gaussian fit and its parameters are plotted in light blue.}
    \label{fig:alphahist}
\end{figure}

 \begin{figure*}[]
	\centering
	\plotone{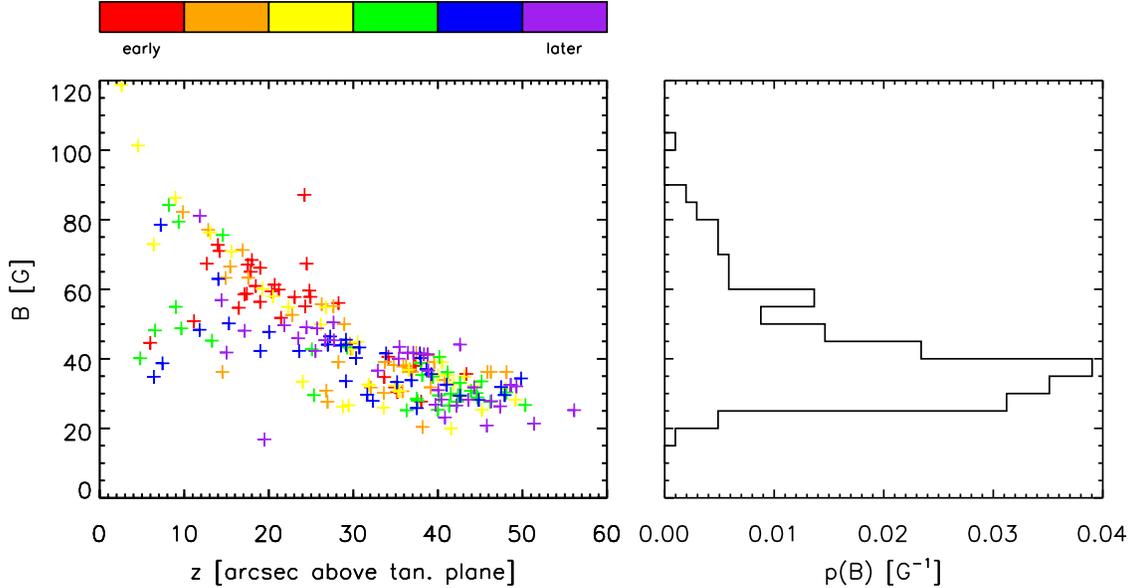}
	\caption{This combination scatterplot and histogram details the magnetic field strengths we obtain from our LFFF model.  Crosses denote B strength vs height and colors correspond to different time intervals.  The intervals were chosen to contain roughly the same number of loops; each color is a group of 33 loops (with red containing 34).\\  The grey histogram shows the B values for all the loops obtained through our LFFF model.  Bins are 5G.}
	\label{fig:bhist}
\end{figure*}

It is possible to compute the flux more accurately than in Equation (\ref{eq:flux}) which assumes that a single B value applies to every single loop at the location where it crosses the slit in this projection.  If each coronal loop traces out a magnetic field line, each should have its own field strength as it crosses the slit.  By using a more sophisticated modeling method, we can assign each loop identified in the stack plot its own field strength and calculate a more informed value for the interconnecting flux.  The coronal magnetic fields of active regions are believed to be in a force-free state that satisfies
\begin{equation}
\nabla \times \mathbf{B} = \alpha(\mathbf{r}) \mathbf{B}
\label{eq:FFeq}
\end{equation}
\citep{1971SoPh...19...72N}. 
A particular case of a force-free field is a linear force-free field (LFFF), or a constant-$\alpha$ field, in which $\nabla \alpha = 0$.  We use this condition along with $\nabla \cdot \mathbf{B} = 0 $ to transform Equation (\ref{eq:FFeq}) into a Helmholtz equation for $\mathbf{B}$.  We will fit a coronal loop to a field line in a LFFF to determine its three-dimensional structure and properties, using the technique from \cite{malanushenko09}.  However, the superposition of LFFF lines with different values of $\alpha$ is not a LFFF field itself.  Thus, a global field containing these LFFF extrapolations not necessarily a force-free field itself.

The method of \cite{malanushenko09} uses magnetogram data to create a set of volume-filling LFFFs and, from a traced coronal loop in an EUV image, determines the most likely LFFF to which it fits.  For each modeling attempt, the Helmholtz equation for $\mathbf{B}$ is solved with the aid of magnetic boundary data at that time from a LOS magnetogram in the half-space \citep{1977ApJ...212..873C,Lothian1995}.  Constant-$\alpha$ fields are computed for an entire volume for 61 values of $\alpha$ equally spaced within the range [-0.05, 0.05] arcsec$^{-1}$.  A coronal loop is manually traced with a smooth curve over an AIA image.  With this method, we need only trace a portion of the loop and the inclusion of visible footpoints is not required.  The coordinates of the loop in the plane of the sky (POS) are known.  The third coordinate, along the line-of-sight, is unknown.

The coordinate along the line-of-sight (LOS) is determined by using field lines traced from various locations, $h$, along the LOS for various constant-$\alpha$ fields, and comparing their projections to the original coronal loop. The line-of-sight along which the field lines are initiated is selected to cross the center of the traced loop in the POS. For every LFFF, a set of such field lines along the LOS are then projected onto the plane-of-sky. An average distance between these projections and the original POS-loop, $d(h, \alpha)$, is then minimized.  Determining the correct minimization is not intuitive, and is detailed in \cite{malanushenko09}.

This modeling was done at 56 time intervals, shown by the crosses in Figure \ref{fig:fluxvoltage}.  These times were chosen to maximize the number of identified loops modeled using the fewest sample times (as building the volume-filling LFFF magnetic models is computationally expensive).  Further, as the POS tracing and selection of the correct minimization is labor intensive, we chose to prioritize modeling of loops that lasted for 10 minutes or longer.  For each instance in time where the modeling is done, every field of view is cut to the same size.  By doing so we accounted for differences in time and the impact they would have on the model.  The volume-filling field was created in a rectilinear domain whose bottom boundary was a plane tangent to the solar surface.  We will reference z-positions with respect to the height above this tangent plane.  We modeled 185 individual loops of the 205 that were identified in the stack plot from Figure \ref{fig:stack} and also lasted 10 minutes or longer.  Due to the manner in which we selected times in an effort to maximize the loops we modeled, there were occasions where a loop was modeled for more than one instance in time.  Thus there are 199 linear force-free field lines being modeled.

For each time when the LFFF model was constructed, all stack plot-identified loops present at the time and satisfying the lifetime criteria were traced to determine their structure.  One such modeling instance is shown for reference in Figure \ref{fig:3dmodel}.  The value of $\alpha$ along the field line is a result of selecting the correct minimization of $d(h,\alpha)$.  The distribution of these values for all of the loops we modeled is shown in Figure \ref{fig:alphahist}.

Of the 16 loops that were modeled at more than one time, most (but not all) of their subsequent modeling yielded their $\alpha$ trending closer to 0.  This was true even for loops with negative $\alpha$: those values trended less negative.  Furthermore, $\alpha$ and emergence time have a Spearman rank-order correlation $\rho=-0.21$, indicating a correlation at a significance level of $99.8\%$.  A linear fit of this data set had positive values with a negative slope.  Because the values of $\alpha$ skew positive, this trend towards $0$ could show the positive wing of the $\alpha$ distribution trending to an equilibrium near 0.

The three-dimensional magnetic field line corresponding to an observed loop yields a value for its full length, and for the height and magnetic field strength at the point it crosses our slit.  (The full length is used in the estimate of cooling time given in \secref{sec:loopprop}.)  A height vs. field strength scatter plot of the modeled loops is shown in Figure \ref{fig:bhist} where the loops are grouped into different colored time intervals, chosen such that each bin has the same number of loops (34 for the red, 33 for the other colors).  The magnetic field strength values obtained from the model are also shown in Figure \ref{fig:bhist}.  If one loop was alive for multiple instances of modeling, its field strength was found from the average of slit-crossing field strengths.  We used a characteristic field strength of 30 G in \secref{sec:loopprop}, which was near the histogram's peak and thus a reasonable choice for a single characteristic value.  However, the distribution skews to the positive side of the mode and neglecting the skew may have resulted in an underestimation of the flux.

\begin{figure*}[]
\centering
    \plotone{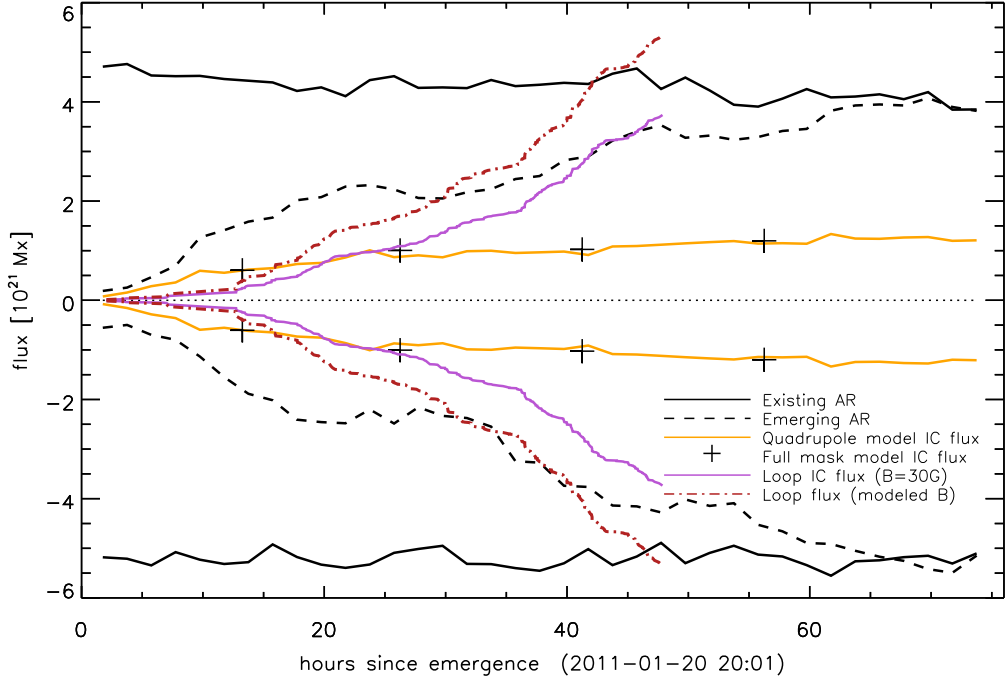}
    \caption{The same flux plot as in figure \ref{fig:potentialflux}, but with the interconnecting loop flux obtained from the loop diameter data from the stack plot with a characteristic (purple) or LFFF modeled (dot-dash red) field strength.}
    \label{fig:withloopfluxfffb}
  \end{figure*}

The interconnecting flux was recalculated using the computed field strength value for each loop (Figure \ref{fig:withloopfluxfffb}, dot-dash red line).  There were 116 loops not modeled individually, which could not be assigned a field strength informed by the LFFF modeling.  To these we assigned the median value B=39 G of the modeled loops.
The interconnecting flux calculated with this method is approximately 50\% greater than when every loop was assigned a common characteristic value of 30 G.  The mean value of field strength of our LFFF-modeled loops at the location of the virtual slit as determined from the LFFF model was 43.7 G, approximately 45\% larger than the characteristic value used in \secref{sec:MCT}, which may play a part in the discrepancy between using the median of the distribution and the mean.  However, not only is the interconnecting flux determined from our stack plot in conjunction with our improved field modeling still larger than the maximum allowed by the MCT model, but the magnitude of interconnecting flux is comparable to the total emerged flux of the newly emerged AR and even surpasses the magnetic flux of the stronger, existing AR's polarities.  This implies that more flux is reconnecting from the emerging positive polarity to the existing active region than is there to begin with. If this were true, then there is an additional source of flux that contributes to this interconnecting region that is not either of the two ARs.  Therefore, more investigation is necessary.

\section{MCT model with non-zero twist}\label{sec:nonzeroalpha}

\begin{figure}[!t]
\centering
    \plotone{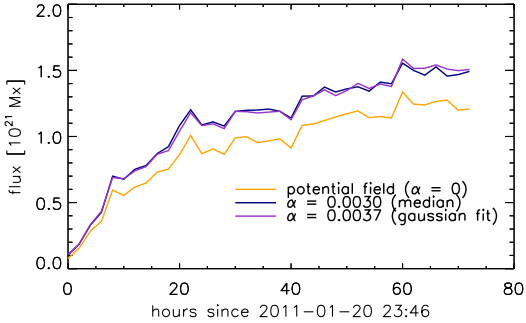}
    \caption{The interconnecting flux of the potential field (orange) is the same as Figure \ref{fig:potentialflux}.  We also repeated this analysis for $\alpha$ values that were the mean (navy) and gaussian (mode) fit (purple) of the distribution in Figure \ref{fig:alphahist}.}
    \label{fig:nonzeroalpha}
\end{figure}

In an effort to find a better coronal magnetic model with which to compare the interconnecting flux, we revisit our initial assumption of a potential field.  The updated flux calculation uses values of $\alpha$ informed by the LFFF modeling.  In the modeling done in \secref{sec:MCT}, the potential field used had an $\alpha=0$ by definition.  We find the interconnecting flux is larger for a non-zero $\alpha$ value compared with the potential field case (where $\alpha =0$, see the orange line in Figures \ref{fig:potentialflux}, \ref{fig:fluxvoltage}, \ref{fig:withloopfluxfffb} and \ref{fig:nonzeroalpha}).
However, we find that this increase is $\approx 20\%$ and does not account for the large discrepancy between the interconnecting fluxes calculated in \secref{sec:loopprop} and \secref{sec:annymodel} (purple and dot-dash red lines in figure \ref{fig:withloopfluxfffb}) when compared to that from the potential field.

To perform this computation we use values of $\alpha$ obtained from the modeling in \secref{sec:annymodel}, which definitely skew positive (see Figure \ref{fig:alphahist}).  This positive skew in values of $\alpha$ may be responsible for the discrepancy between the high interconnecting flux we see in contrast to what a potential field allows.  The non-potentiality of the field, suggested by the modeled loops, might partition the flux in a different manner such that there is more flux allowed in the interconnecting region than when compared to the lowest energy state of the field.  We redid the interconnecting flux calculation using a non-zero $\alpha$ in the MCT model (these values are detailed below).  The same quadrupole models from \secref{sec:MCT} were used to perform the extrapolation for these new field and subsequent interconnecting flux, using the Green's function from \cite{1977ApJ...212..873C}.  Unlike the modeling in the previous section, the global field in the MCT model detailed here is a LFFF.

Two extrapolations were done to calculate interconnecting flux in a LFFF model, for two different values of $\alpha$: one corresponding to the peak of a Gaussian fit to the histogram in Figure \ref{fig:alphahist} ($\alpha = 3.7 \times 10^{-3}$ arcsec$^{-1}$), and the second one corresponding to the median of the same distribution ($\alpha = 3.0 \times 10^{-3}$ arcsec$^{-1}$).  The results of these are shown in purple and navy, respectively, with the original $\alpha=0$ results in orange on Figure \ref{fig:nonzeroalpha}.  There is an increase in the magnitude of interconnecting flux between the two ARs for a non-zero twist, however it is nowhere large enough to account for the results determined from the stack plot with AIA data.

\section{Loop Locations in the Plane Defined by the Slit and the Line of Sight}\label{sec:slitplane}

\begin{figure*}
\centering
\plotone{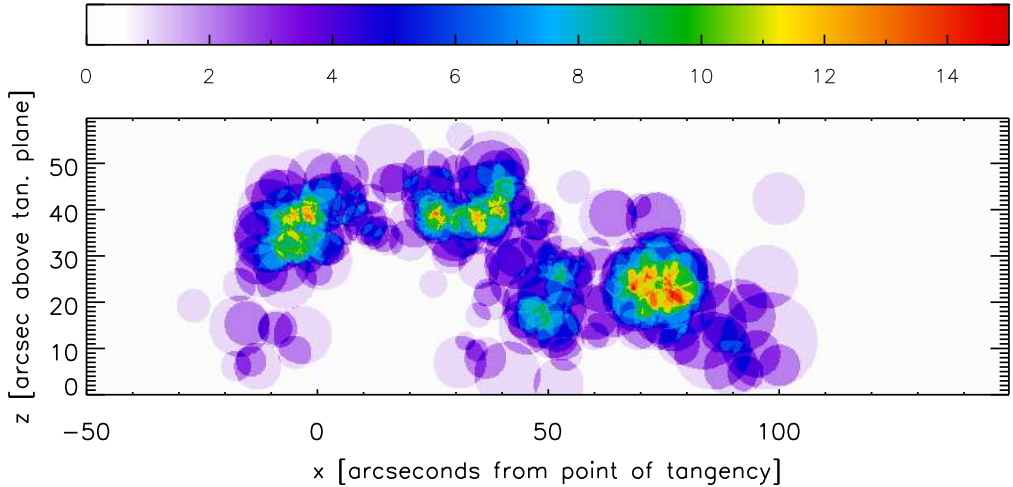}
    \caption{A 2D histogram showing the locations of loops in the plane which projects into the virtual slit.  There appear to be distinct ``hot-spots'' where many loop cross-sections lie on top of each other and ``pockets'' where no loops appear to be. The maximum value is 15 loop cross-sections lying on one another in 48 hours of data, or $8.7 \times 10^{-5}$ Hz.}
    \label{fig:planerate}
\end{figure*}

\begin{figure}[]
\centering
    \plotone{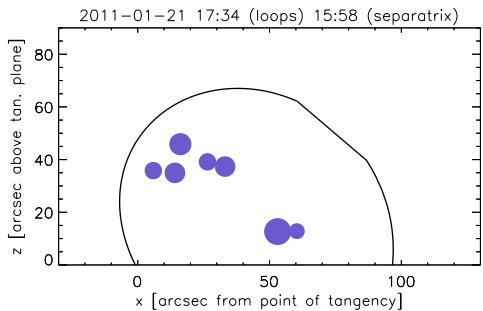}
    \caption{The intersection of the virtual slit with the separatrix (MCT model) and loops (LFFF model) from similar times.  As in Figure \ref{fig:planerate}, the loops are clustering in ``hot spots''.}
    \label{fig:seprover}
\end{figure}

\begin{figure*}[]
\centering
    \plotone{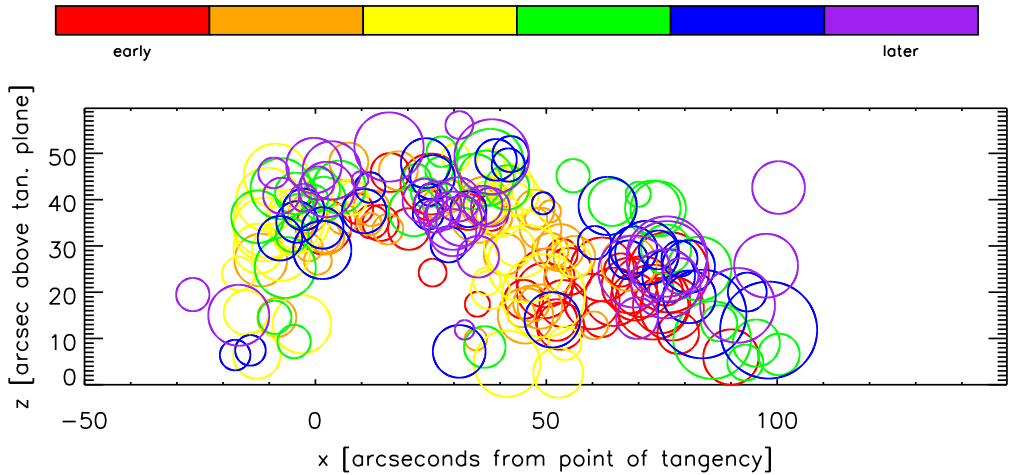}
    \caption{A version of Figure \ref{fig:planerate}, wherein the time evolution is depicted by the different colored bin.  The loops are binned in the same manner as in figure \ref{fig:bhist}.  An animated version of this figure is available.}
    \label{fig:figure15}
\end{figure*}

With the benefit of three-dimensional loop information we see that it is likely that some flux has brightened repeatedly during our 48-hour observation.
Figure \ref{fig:planerate} shows a two-dimensional histogram depicting the location of the modeled loops in the plane defined by two vectors: along the LOS and along the slit. We assume a cylindrical loop, and that each cylinder intersects the plane of the virtual slit normally, so each intersection is a circle.  The center of this disk lies at the position in the tangent plane determined from the LFFF modeling in \secref{sec:annymodel}.  The pixel color denotes the number of different disks that overlap at that point.  The loop cross-sections lie on top of each other in the xz-plane, with three distinct locational hot-spots and at least one empty pocket where there appears to be no loops.  The most apparent empty pocket to which we refer is in the range 0-50'' from the point of tangency, and at a height of less than 30'' above the tangent plane.  Figure \ref{fig:seprover} shows a selection of modeled coronal loops from \secref{sec:annymodel} with the separator from a similar time from the MCT model in \secref{sec:MCT}.  Figure \ref{fig:figure15} depicts a version of Figure \ref{fig:planerate}, showing the time evolution trends of the loops in that plane.  (An animated version of Figure \ref{fig:figure15} is available.)  The discrete color bins used in Figure \ref{fig:figure15} are the same as those used in the left panel of Figure \ref{fig:bhist}.

Due to the manner in which our interconnecting loops lie on top of each other in the plane of the virtual slit, it may be that our assumption that a coronal loop brightens only when it is newly reconnected is faulty.  Perhaps not all brightening interconnecting loops are independent samplings of new flux and in assuming so we have overestimated substantially the flux linkage between the two active regions.  Under this assumption we neglected to consider any internal reconnection within the magnetic domain where the interconnecting loops are located.  Loops in that domain reconnecting to achieve a lower energy state would be brightening loops that cross our virtual slit, but would not be newly reconnected flux from between the two AR.

Our data suggests that we are seeing a flux element brighten multiple times.  
We calculate the minimum amount of reconnected flux supported from our analysis by first considering an overlap percentage obtained from the 2D-histogram in Figure \ref{fig:planerate}.  This percentage is that of the pixels with counts of more than 1 in the histogram, over the total number of pixels with nonzero counts.  This yields a percent overlap of $70.77\%$.  Applying the likelihood of non-overlap ($29.23\%$) to the maximum interconnecting flux value from Figure \ref{fig:withloopfluxfffb} (red, dot-dash line) of $5.35\times10^{21}$ Mx returns a scaled value that, to first order, corrects for overlap we initially attributed to independent samplings of flux.  This scaled value is $1.56 \times 10^{21}$ Mx, which is much closer to the interconnecting flux predicted by the MCT models.

\section{Discussion and conclusions}\label{sec:dis}

Due to the advent of AIA and the subsequent increased availability of observations of emerging/existing AR pairs, we perform analysis to quantify the reconnection during one such emergence event between two distinct magnetic systems.  We find we overestimate the amount of interconnecting flux we identify, despite potential undersampling of the loops with footpoints anchored in opposite ARs due to our decision to not include loops with short lifetimes in our analysis. We assert that this may be due to seeing multiple brightenings of a single magnetic flux element throughout our observations \citep{Webb1981}.

We propose that the repeated brightening of loops in similar spatial locations are the result of Taylor relaxation of the coronal field within the domain which connects the emerging positive polarity to the existing negative \citep{1986RvMP...58..741T,Nandy2005}.  Reconnection occurs within the domain in order to reach a lower energy state.  The decay of a nonpotential coronal field has been observed on timescales of tens of hours after new flux emerges \citep{2005ApJ...628..501S}.  We assert that we see the field relaxing to one with constant $\alpha$  after an interconnecting loop is created through reconnection.  As we see a distribution of $\alpha$-values in our LFFF modeling of individual coronal loops, we know that the field between the two ARs is not in its lowest energy states.  There is ample time for internal reconnection of the interconnecting loops due to the slower pace of reconnection and flux evolution driven by the rate of emergence of AR11149 and the resulting topological evolution of the two AR system.

The result of the large interconnecting flux inferred from the data lead to an artificially high reconnection rate.  We found a rate of $2.13 \times 10^{16}$ Mx s$^{-1}$ (yielding a total reconnection time of 2.72 days when dividing the strength of the existing AR of $\sim5\times10^{21}$ Mx by this rate) over two days.  \cite{Tarr2014} examined quiescent reconnection within single NOAA AR11112 and inferred a reconnection rate of $0.38 \times 10^{16}$ Mx s$^{-1}$ over the same time interval.  This is a reconnection time of $\sim16.74$ days for an AR of similar strength, where our inflated reconnection time value is just $~16\%$ in comparison.  And though the conclusions of \cite{2005ApJ...630..596L} was the underestimation of the reconnection rate, their inferred rate was $3.97 \times 10^{16}$ Mx s$^{-1}$ ($3.49$ days), though over a substantially smaller time window of approx. 3.5 hours wherein there was an abundance of reconnection.  The reconnection rate in this study peaked at approx. $7 \times 10^{16}$ Mx s$^{-1}$ ($0.83$ days), or $\sim175\%$ larger than the reconnection rate of \cite{2005ApJ...630..596L}.  The median diameter of our loop distribution $5.9$ Mm is larger than the $3.7$ Mm result of \cite{2005ApJ...630..596L}, a potential result of our boxcar smoothing during our peak finding procedure.  Though these enhanced diameters could contribute to the excess of flux we see, we do not believe this is the source of significant over-counting of the interconnecting flux as our median diameter result is comparable to those of other studies with maximum diameters of $\sim 5.6-6$ Mm with both TRACE and AIA \citep{2000ApJ...541.1059A,2013SoPh..283....5A}.

Of particular note between this work and that of \cite{2005ApJ...630..596L}, though the properties of loops obtained in both are similar, the discrepancy in number of loops cataloged (43 vs 301 in this work) may point to a significant topological difference between the two studies.  MHD simulations have explored the parameter space regarding the favorability of an AR to reconnect with a neighbor.  Such a difference in the amount of observed loops could be a result of the physical characterization of the polarities between these two regions \citep{2017ApJ...850...39T}, or whether the field lines of opposite ARs are oriented parallel or anti-parallel to each other \citep{Galsgaard_2007}.  This topological difference could manifest in many ways, including in the delay between an AR emergence and the onset of reconnection to the overlying field.  The observed delay in reconnection, both in this study and elsewhere in the literature \citep{2005ApJ...630..596L,2008A&A...488.1117Z,KobelskiThesis}, might be a result of how inclined the fields are to reconnect given their $\alpha$ values \citep[as in][in regards to co- and counterhelicity]{Linton2001,Yamada1990}.  The evolution of the AR system may reach some critical point, allowing for reconnection that changes a property of the flux tube (like $\alpha$) instead of the exchange of new flux between domains.  The change in twist may cause some sort of cascade effect, wherein Taylor-like reconnection is able to take the field to an even lower energy state.  

We aim to refine the methods detailed herein to improve the reliability of our loop catalog through a more a robust loop identification and selection technique.  This is especially vital for a more thorough investigation of those loops with shorter lifetimes below our threshold in this work.  Applying the improved method to a wider set of emerging/existing AR pairs will allow us to quantify what considerations, particularly with cohelicity/counterhelicity between the two, play a role in this kind of coronal reconnection process.

The foregoing analysis has assumed the coronal loops observed in EUV had circular cross sections.  While this is a fairly common assumption, it can be called into question \citep{2013ApJ...775..120M}.  Doing so might offer additional insight into, or an alternative explanation for, the large flux values we have obtained.

\bigskip
This work was supported by NASA's HGI program.

\bigskip
We would like to thank the anonymous referee for his or her helpful suggestions for improving both this manuscript as well as the future work involving emerging/existing active region pair reconnection.

\bibliography{paperbib}
\bibliographystyle{aasjournal}

\end{document}